\documentclass{elsart}
\usepackage{graphicx}
\journal{Physica D}
\catcode`\@=11
\newcommand{\beq}{\begin{equation}}
\newcommand{\eeq}{\end{equation}}
\newcommand{\abs}[1]{\vert#1\vert}
\newcommand{\absln}[1]{\vert\!\ln#1\vert}

\renewcommand{\d}{{\rm d}}

\newcommand{\eps}{\varepsilon}
\renewcommand{\flat}{{\rm flat}}
\renewcommand{\frac}[2]{\displaystyle{\displaystyle#1\over\displaystyle#2}}
\newcommand{\mean}[1]{\langle#1\rangle}

\newcommand{\s}{\sigma}
\newcommand{\sign}{\mathop{\rm sign}\nolimits}
\newcommand{\xidy}{\xi_{\rm dyn}}
\newcommand{\C}{{\mathcal C}}
\newcommand{\Frac}{\mathop{\rm Frac}\nolimits}
\newcommand{\Int}{\mathop{\rm Int}\nolimits}
\newcommand{\N}{{\mathcal N}}

\def\numberbysection{\@addtoreset{equation}{section}
\def\theequation{\thesection.\arabic{equation}}}
\numberbysection

\begin{document}
\begin{frontmatter}

\title{Ground-state structure and dynamics
in a toy~model for granular compaction}

\author{J.M.~Luck}

\ead{luck@spht.saclay.cea.fr}

\address{Service de Physique Th\'eorique\thanksref{cnrs},
CEA Saclay, 91191~Gif-sur-Yvette cedex, France}

\thanks[cnrs]{URA 2306 of CNRS}

\begin{abstract}
We report on a toy model
for the glassy compaction dynamics of granular systems,
introduced and investigated in collaboration with Anita Mehta and Peter Stadler.
A stochastic dynamics is defined on a column of grains.
Grains are anisotropic and possess a discrete orientational degree of freedom.
Gravity induces long-range directional interactions down the column.
The key control parameter of the model, $\eps$,
is a representation of granular shape.
Rational and irrational values of $\eps$ correspond to very different
kinds of behavior, both in statics (structure of ground states)
and in low-temperature dynamics (retrieval of ground states).
\end{abstract}

\end{frontmatter}

\section{Foreword}

In this contribution I have chosen to report on a series of works
done in recent years in collaboration
with Anita Mehta and Peter Stadler~\cite{I,II,III,IV},
where we have introduced and investigated in detail
a columnar {\it toy model}
for the compaction dynamics of granular systems in the glassy regime.
The main focus of this review is on the properties of this model
which are of interest in the broader context of
non-equilibrium dynamics and complex systems.
We emphasize the relationship between statics (structure of ground states)
and low-temperature dynamics (retrieval of these ground states).
Surprisingly enough,
some of the key features of this model belong to Serge Aubry's favorites,
such as {\it quasiperiodicity} and {\it hierarchical melting}.

\section{Definition of the model}

A columnar model for the compaction dynamics of granular media
in the glassy regime has been elaborated and investigated
in a series of works~\cite{I,II,III,IV}.
The model consists in a column of $N$ sites,
each site being occupied by a grain.
Grains are assumed to be anisotropic in shape.
They take, for simplicity, two possible orientations,
referred to as {\it ordered} and {\it disordered}.
Sites are labeled by their depth $n=1,\dots,N$,
measured from the top of the column.
Orientation variables are defined
by setting $\s_n=+1$ if grain number~$n$ is ordered,
and $\s_n=-1$ if grain number $n$ is disordered.
A configuration of the system is thus defined
by the $N$ orientation variables $\{\s_n\}$.

The collective dynamics of the orientation variables is chosen
in order to model the compaction dynamics of granular systems
under the influence of a small vibration intensity.
The crucial effect of gravity is taken into account by
considering directional long-range interactions.
Grain number $n$ feels the effective weight of the whole piece
of column above it ($m=1,\dots,n-1$).
As a consequence,
the upper grain $(m)$ influences the dynamics of the lower one $(n)$,
whereas the lower grain has no influence on the dynamics of the upper one.
In other words, causality acts both in space and in time.
Hence there are no finite-size effects:
a finite system made of $N$ grains behaves exactly as the $N$ upper grains
of a larger (finite or infinite) system.

More specifically, we consider a Markovian stochastic dynamics
defined by the following transition rates per unit time:
\beq
\left\{\matrix{
w(\s_n=+1\to\s_n=-1)=\exp\left(-\frac{\lambda_n+h_n}{T}\right),\hfill\cr\cr
w(\s_n=-1\to\s_n=+1)=\exp\left(-\frac{\lambda_n-h_n}{T}\right),\hfill}\right.
\label{w}
\eeq
where $T$ is a dimensionless measure of the vibration intensity,
referred to as temperature,
whereas the effective activation energy $\lambda_n$
and the effective ordering field $h_n$
represent the effect of all the grains above grain number~$n$.

In the most general model, $\lambda_n$ and $h_n$ depend
on all the orientation variables~$\s_m$ for $m=1,\dots,n-1$.
The simplest non-trivial model corresponds to~\cite{III,IV}
\beq
\lambda_n=A\,n,\qquad h_n=\eps\,m^-_n-m^+_n,
\label{ydef}
\eeq
where $m^+_n$ (resp.~$m^-_n$)
is the number of ordered (resp.~disordered) grains above grain number~$n$:
\beq
m^+_n=\frac12\sum_{k=1}^{n-1}(1+\s_k),\qquad
m^-_n=\frac12\sum_{k=1}^{n-1}(1-\s_k),
\eeq
so that $m^+_n+m^-_n=n-1$.

The shape parameter $\eps$ is the essential control parameter of the model.
It has to obey $\eps>0$ in order for the model to be non-trivial.
It will be identified with the slope involved in
the geometrical construction given below.
Rational and irrational values of $\eps$,
respectively corresponding to regular and irregular grain shapes,
turn out to correspond to very different kinds of static and dynamical behavior.

The amplitude $A$ of the effective activation energy
gives rise to a dynamical length
\beq
\xidy=\frac{T}{A},
\label{xidydef}
\eeq
such that the Arrhenius factor of the rates at depth~$n$ reads
\beq
\omega_n=\exp\left(-\frac{\lambda_n}{T}\right)
=\exp\left(-\frac{n}{\xidy}\right).
\label{omeg}
\eeq

\section{Ground-state structure}

As the dynamical rules of the model are fully directional,
they cannot obey detailed balance with respect to any Hamiltonian.

There is, however, a well-defined concept of {\it ground state}.
In the zero-tem\-pe\-ra\-ture limit we have indeed
\beq
\frac{w(\s_n=-1\to\s_n=+1)}{w(\s_n=+1\to\s_n=-1)}
=\exp\left(\frac{2h_n}{T}\right)\to\left\{\matrix{
\infty\hfill&\hbox{if}\hfill&h_n>0,\cr
0\hfill&\hbox{if}\hfill&h_n<0.
}\right.
\label{zero}
\eeq
We are therefore led to define a ground state of the system
as a configuration where the orientation of every grain
is aligned along its local field:
\beq
\s_n=\sign h_n=\left\{\matrix{
+1\hfill&\hbox{if}\hfill& h_n>0,\cr
-1\hfill&\hbox{if}\hfill& h_n<0\hfill
}\right.
\label{zerost}
\eeq
(provided $h_n\ne0$).
The condition~(\ref{zerost})
leaves the uppermost orientation~$\s_1$ unspecified,
as $h_1$ vanishes identically.
In the following, we assume for definiteness
that the uppermost grain is ordered:
\beq
\s_1=+1.
\label{init}
\eeq

\begin{table}
\caption{Recursive construction of ground states by means of~(\ref{zerost}).}
\smallskip
\begin{tabular}{|c|c|c|c|c|}
\hline
$h_n$&$\s_n$&$m^+_{n+1}-m^+_n$&$m^-_{n+1}-m^-_n$&$h_{n+1}-h_n$\\
\hline
$>0$&$+1$&$1$&$0$&$-1$\\
$<0$&$-1$&$0$&$1$&$\eps$\\
$=0$&?&?&?&?\\
\hline
\end{tabular}
\label{hphn}
\end{table}

Ground states can be built in a recursive way~(see Table~\ref{hphn}).
The number and the nature of ground states depend
on whether $\eps$ is irrational (only~$\s_1$ is unspecified)
or rational (infinitely many orientations are unspecified).
These two situations will be considered successively.

\subsection{Irrational $\eps$: Unique quasiperiodic ground state}

For irrational $\eps$, all the local fields $h_n$ are non-zero.
Table~\ref{hphn} implies that they lie in the bounded interval
\beq
-1\le h_n\le\eps.
\label{strip}
\eeq

\begin{figure}
\begin{center}
\includegraphics[angle=90,width=.75\linewidth]{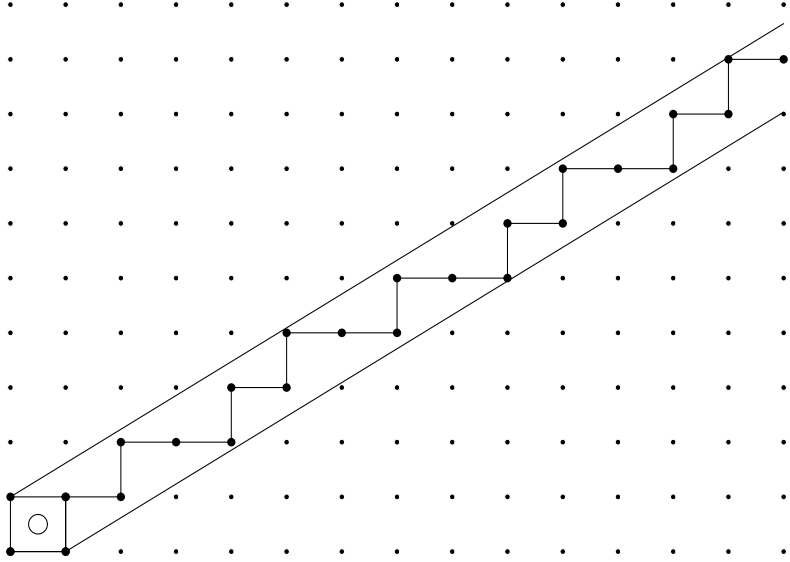}
\end{center}
\caption{
Geometrical construction of the quasiperiodic ground state of the model
for the golden-mean slope~(\ref{gold}).
The two ways of going around the first cell, marked with a circle,
correspond to the two possible choices for the orientation
of the uppermost grain (after~\cite{IV}).}
\label{figa}
\end{figure}

Let us introduce the following {\it superspace formalism}.
Consider the integers $(m^-_n,m^+_n)$ as the co-ordinates
of points on a square lattice.
We thus obtain a staircase-shaped line joining the points
$(m^-_1,m^+_1)=(0,0)$, $(m^-_2,m^+_2)=(0,1)$, etc.
Vertical steps correspond to ordered grains,
whereas horizontal steps correspond to disordered grains.
Equation~(\ref{strip}) defines an oblique strip with {\it slope}~$\eps$
in the $(m^-,m^+)$ plane,
which contains the entire line thus constructed~(see Figure~\ref{figa}).

A unique infinite ground-state configuration of grain orientations
is thus generated.
This configuration is {\it quasiperiodic}.
The above construction is indeed equivalent to the cut-and-project method
of generating quasiperiodic tilings of the line.
This approach,
introduced by de Bruijn in the mathematical literature~\cite{deB},
became then very popular in the context of quasicrystals~\cite{quasi}.

The ground state thus constructed can alternatively be described
in analytical terms.
We have indeed
\beq
m^+_n=n-1-m^-_n=1+\Int((n-1)\Omega),\qquad h_n=g(n\Omega),
\label{rot}
\eeq
where the {\it rotation number} $\Omega$ reads
\beq
\Omega=\frac{\eps}{1+\eps}
\eeq
and the {\it hull function} $g(x)$ is a $1$-periodic discontinuous
sawtooth function:
\beq
g(x)=-1+\frac{\Frac(x-\Omega)}{1-\Omega}.
\eeq
In these formulas $\Int(x)$, the integer part of a real number $x$,
is the largest integer less than or equal to $x$,
whereas $\Frac(x)=x-\Int(x)$ is the fractional part of $x$ $(0\le\Frac(x)<1)$.

There are well-defined proportions of ordered and disordered grains
in the ground state:
\beq
f_+=\Omega=\frac{\eps}{1+\eps},\qquad f_-=1-\Omega=\frac{1}{1+\eps}.
\label{propor}
\eeq

This geometrical construction is illustrated in Figure~\ref{figa}
for the most familiar of all irrational numbers~\cite{hr},
the inverse golden mean:
\beq
\eps=\Phi-1=\frac{1}{\Phi},\quad\Omega=2-\Phi=\frac{1}{\Phi^2},\quad
\Phi=\frac{\sqrt5+1}{2}\approx1.618033.
\label{gold}
\eeq
The corresponding configuration is given by a Fibonacci
sequence~\cite{deB,quasi,hr}:
\[
\{\s_n\}=+--+--+-+--+--+-+--+-+--\cdots
\]

\subsection{Rational $\eps$: Degenerate ground states}

For a rational slope
\beq
\eps=\frac{p}{q},\qquad\hbox{i.e.,}\qquad\Omega=\frac{p}{p+q},
\eeq
in irreducible forms ($p$ and $q$ are mutual primes),
some of the local fields $h_n$ generated by Table~\ref{hphn} vanish.
The corresponding grain orientations $\s_n$ remain unspecified.

\begin{figure}
\begin{center}
\includegraphics[angle=90,width=.75\linewidth]{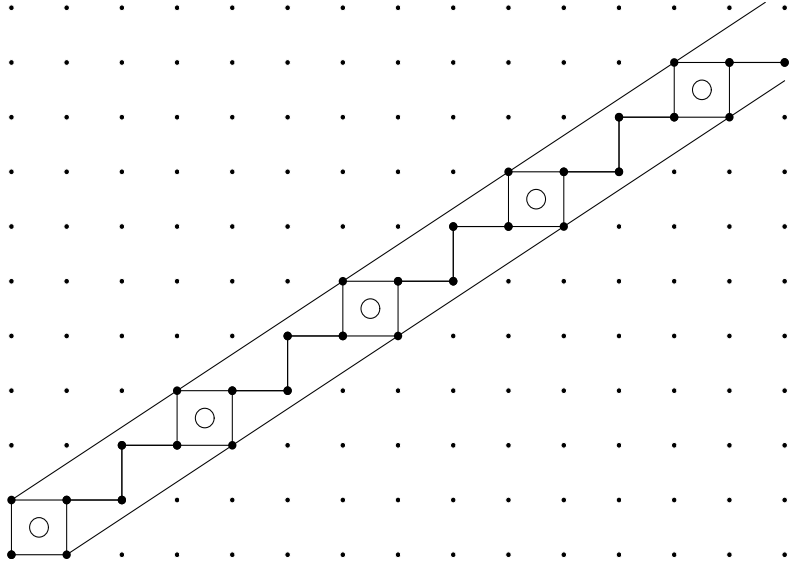}
\end{center}
\caption{
Geometrical construction of the ground states of the model
for the rational slope $\eps=2/3$.
The marked cells, entirely contained in the strip,
are responsible for the non-zero configurational entropy (after~\cite{IV}).}
\label{figb}
\end{figure}

This feature of rational slopes is clearly visible
on the geometrical construction.
Figure~\ref{figb}, corresponding to $\eps=2/3$,
shows that some of the lattice cells, marked with circles,
are entirely contained in the closed strip~(\ref{strip}).
Consider one such cell.
The broken line enters the cell at its lower left corner
and exits the cell at its upper right corner.
It can go either counterclockwise, via the lower right corner,
giving $\s_{n+1}=-1$, $\s_{n+2}=+1$,
or clockwise, via the upper left corner, giving $\s_{n+1}=+1$, $\s_{n+2}=-1$.
Each marked cell thus generates a binary choice in the construction.
This orientational indeterminacy occurs at points
such that~$n$ is a multiple of the {\it period} $p+q$,
equal to the denominator of the rotation number $\Omega$.

The model therefore has a non-zero ground-state entropy
(zero-temperature configurational entropy, complexity~\cite{sconf})
\beq
\Sigma=\frac{\ln 2}{p+q}
\label{sigma}
\eeq
per grain.
Each ground state is a random sequence of two
well-defined patterns of length $p+q$,
each of them made of $p$ ordered and $q$ disordered ones,
so that~(\ref{propor}) still holds for each of the ground states.
The patterns only differ by their first two orientations.
The first cases are listed in Table~\ref{tab}.

\begin{table}
\caption{
Patterns building up the random ground states
for the first rational values of $\eps$.
The second example with period $p+q=5$ is illustrated in Figure~\ref{figb}
(after~\cite{IV}).}
\smallskip
\begin{tabular}{|c|c|c|c|c|c|c|}
\hline
$p+q$&$\Omega$&$\eps$&$p$&$q$&pattern 1&pattern 2\\
\hline
2&1/2&1&1&1&$+-{}$&$-+{}$\\
\hline
3&1/3&1/2&1&2&$+--$&$-+-$\\
3&2/3&2&2&1&$+-+$&$-++$\\
\hline
4&1/4&1/3&1&3&$+---{}$&$-+--{}$\\
4&3/4&3&3&1&$+-++{}$&$-+++{}$\\
\hline
5&1/5&1/4&1&4&$+----$&$-+---$\\
5&2/5&2/3&2&3&$+--+-$&$-+-+-$\\
5&3/5&3/2&3&2&$+-+-+$&$-++-+$\\
5&4/5&4&4&1&$+-+++$&$-++++$\\
\hline
6&1/6&1/5&1&5&$+-----{}$&$-+----{}$\\
6&5/6&5&5&1&$+-++++{}$&$-+++++{}$\\
\hline
\end{tabular}
\label{tab}
\end{table}

\section{Zero-temperature dynamics.~Retrieval of ground states}

We have already anticipated in the previous section
that the stochastic dynamics of the model simplifies
in the zero-temperature limit.
Equation~(\ref{zero}) indeed becomes the deterministic rule
\beq
\s_n\to\sign h_n
\label{zerody}
\eeq
(again provided $h_n\ne0$).

The dynamical length $\xidy$ introduced in~(\ref{xidydef})
is assumed to have a non-trivial zero-temperature limit.
Most of the interesting features of the model are already present in the
limiting case where $\xidy=\infty$.

\subsection{Irrational $\eps$, infinite $\xidy$: Ballistic coarsening}

For irrational $\eps$, the rule~(\ref{zerody}) is always well-defined,
as the local fields $h_n$ never vanish.
We assume that the system is initially in a disordered state,
where each grain is oriented at random: $\s_n=\pm$ with equal probabilities,
except for the uppermost one, which is fixed according to~(\ref{init}).

Consider first the limiting situation where $\xidy$ is infinite.
The zero-tem\-pe\-ra\-ture dynamics is observed to drive
the system to its quasiperiodic ground state.
This ordering propagates down the system from its top,
via {\it ballistic coarsening}.
At time $t$, the grain orientations have converged
to their ground-state values, given by the above geometrical construction,
in an upper layer whose depth is observed to grow linearly with time:
\beq
L(t)\approx Vt,
\label{vt}
\eeq
whereas the rest of the system is still disordered.
This phenomenon is similar to phase ordering,
as order propagates over a macroscopic length $L(t)$ which grows forever.
It is however different from usual coarsening,
as the depth of the ordered region grows ballistically,
with a well-defined $\eps$-dependent ordering velocity $V(\eps)$,
instead of diffusively, or even more slowly~\cite{bray}.

\begin{figure}
\begin{center}
\includegraphics[angle=90,width=.75\linewidth]{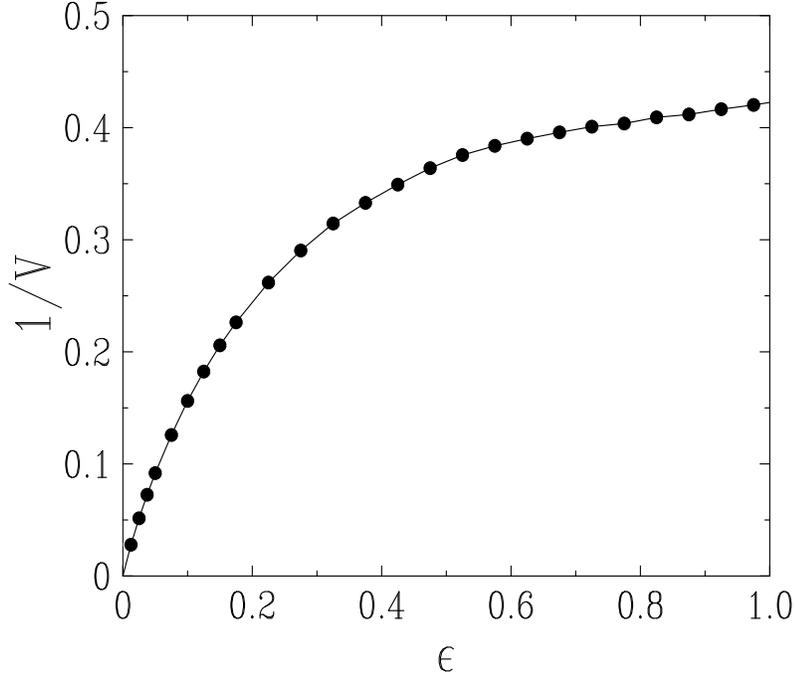}
\end{center}
\caption{
Plot of the inverse ordering velocity $1/V$
of zero-temperature coarsening dynamics at infinite $\xidy$,
against the irrational slope $\eps$, for $0<\eps<1$.}
\label{figc}
\end{figure}

The ordering velocity obeys the symmetry property $V(\eps)=V(1/\eps)$.
Besides this property, its $\eps$-dependence can only
be investigated numerically.
The ordering velocity is observed to vary smoothly with~$\eps$
(although it is only defined for irrational~$\eps$),
and to diverge as $V(\eps)\sim1/\eps$ as~$\eps\to0$.
Figure~\ref{figc} shows a plot of the inverse ordering velocity
against $\eps$, for $0<\eps<1$.

\subsection{Irrational $\eps$, finite $\xidy$:
Crossover to logarithmic coarsening}

In the general situation where~$\xidy$ is finite,
but large at the microscopic scale of a grain,
the ballistic coarsening law~(\ref{vt}) is modified in a minimal way,
by taking into account the dependence of the rate~(\ref{omeg}) on depth:
$\d L/\d t\approx V\,\exp(-L/\xidy)$.
We thus obtain
\beq
L(t)\approx\xidy\,\ln\left(1+\frac{Vt}{\xidy}\right).
\label{cross}
\eeq
This result exhibits a crossover between
the ballistic law~(\ref{vt}) for $Vt\ll\xidy$
and the logarithmic coarsening law
\beq
L(t)\approx\xidy\,\ln t
\eeq
for $Vt\gg\xidy$.

\begin{figure}
\begin{center}
\includegraphics[angle=90,width=.75\linewidth]{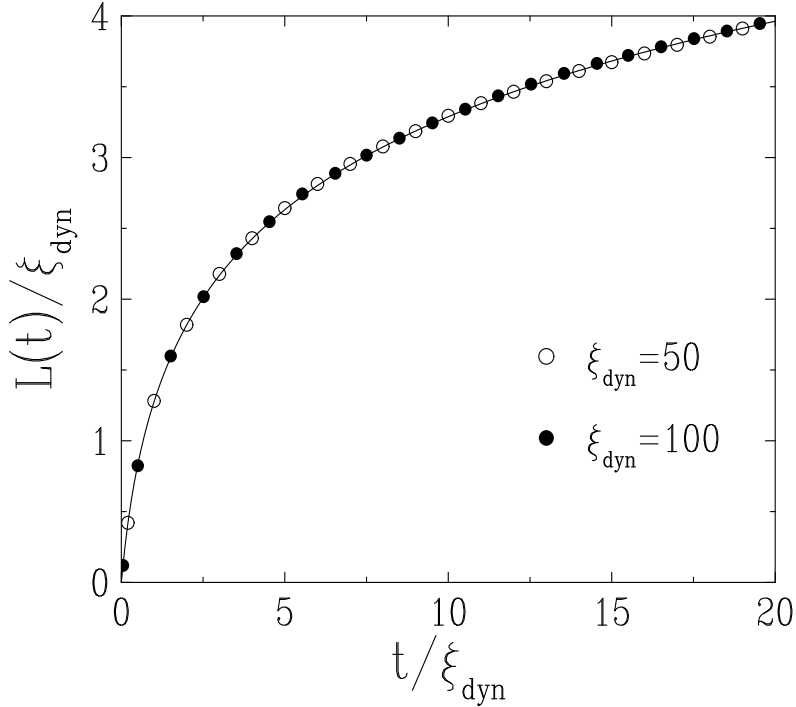}
\end{center}
\caption{
Scaling plot of $L(t)/\xidy$ against $t/\xidy$
for zero-temperature coarsening dynamics with the golden-mean slope.
Symbols: numerical data.
Full line: prediction~(\ref{cross}), with $V=2.58$.}
\label{figd}
\end{figure}

Equation~(\ref{cross}) has been checked
against the results of numerical simulations,
for the golden-mean slope.
Figure~\ref{figd} shows a scaling plot of $L(t)$
corresponding to $\xidy=50$ and 100, together with the
prediction~(\ref{cross}), with no adjustable parameter.
The ordering velocity $V\approx2.58$
is taken from the data of Figure~\ref{figc}.

\subsection{Rational $\eps$: Fluctuating steady state with anomalous roughening}

We now turn to zero-temperature dynamics for rational $\eps$.
The updating rule~(\ref{zerody}) is not always well-defined,
as the local fields $h_n$ may now vanish.
In such a circumstance, it is natural to choose the corresponding orientation
at random:
\beq
\s_n\to\left\{\matrix{
+1\hfill&\hbox{if}\hfill& h_n>0,\cr
\pm1\hfill\hbox{ with prob.~}1/2\hfill&\hbox{if}\hfill& h_n=0,\cr
-1\hfill&\hbox{if}\hfill& h_n<0.\cr
}\right.
\label{zerodyrat}
\eeq
The zero-temperature dynamics defined in this way
therefore keeps a stochastic component.

For concreteness,
let us focus our attention onto the limiting situation
where $\xidy$ is infinite,
and onto the simplest rational case, i.e., $\eps=1$.
Equation~(\ref{ydef}) for the local fields reads
\beq
h_n=-\sum_{m=1}^{n-1}\s_m.
\eeq

It turns out that the zero-temperature dynamics
is not able to retrieve any of the degenerate ground states.
The system rather reaches a non-trivial
{\it fluctuating non-equilibrium steady state}.
This steady state is unique, i.e., independent of the initial configuration
of the system.
It is reached after a microscopic time.

\begin{figure}
\begin{center}
\includegraphics[angle=90,width=.75\linewidth]{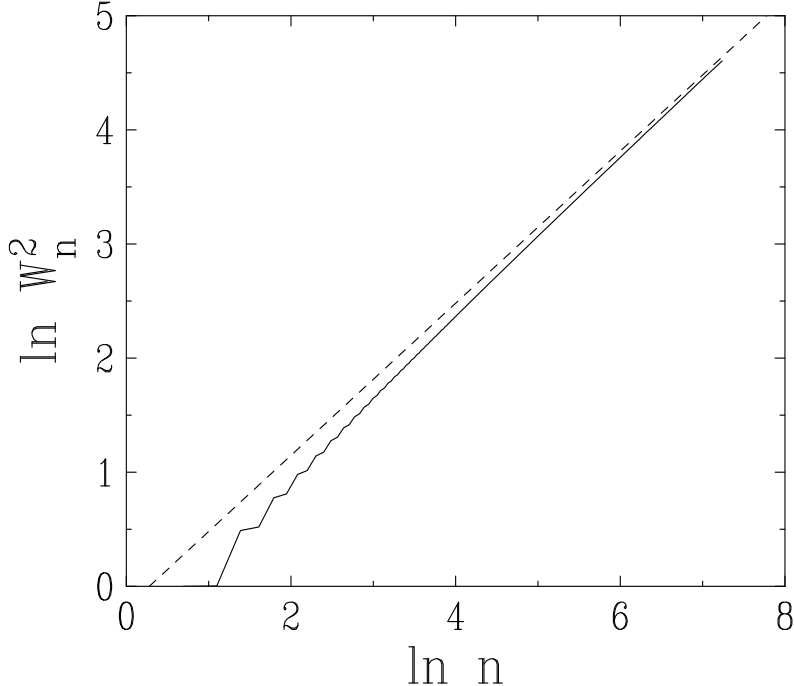}
\end{center}
\caption{
Log-log plot of the local field variance $W_n^2=\mean{h_n^2}$
against depth $n$, for zero-temperature dynamics with $\eps=1$.
Full line: numerical data.
Dashed line: fit to asymptotic behavior, leading to~(\ref{rough})
with $A\approx 0.83$ (after~\cite{III}).}
\label{fign}
\end{figure}

This steady state is characterized by unbounded,
albeit subextensive fluctuations of the local fields.
Figure~\ref{fign} indeed demonstrates that the local field variance grows as
\beq
W_n^2=\mean{ h_n^2}\approx A\,n^{2/3}.
\label{rough}
\eeq

The exponent $2/3$ of this {\it anomalous roughening} law
can be predicted by the following line of reasoning.
Viewing the depth $n$ as time,
and the local field~$h_n$ as the position of a fictitious random walker,
the noise in this walk originates in the earlier times $m$ such that $h_m=0$.
The probability for $h_m$ to vanish is of the order of $1/W_m$.
The effective diffusion coefficient at time~$n$ therefore scales as
$D_n\sim(1/n)\sum_{m=1}^{n-1}1/W_m\sim1/W_n$,
hence $W_n^2\sim nD_n\sim n/W_n$, i.e., $W_n^2\sim n^{2/3}$.

\begin{figure}
\begin{center}
\includegraphics[angle=90,width=.75\linewidth]{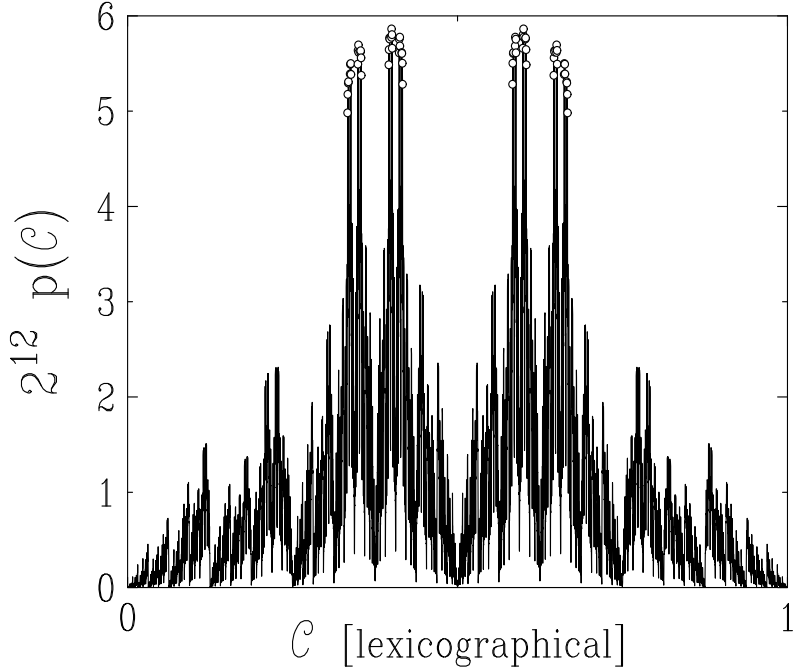}
\end{center}
\caption{
Plot of the normalized probabilities $2^{12}\,p(\C)$ of the configurations
of a column of 12 grains in the zero-temperature steady state with $\eps=1$,
against the configurations $\C$ in lexicographical order.
Open circles: degenerate ground states (after~\cite{IV}).}
\label{figigs}
\end{figure}

Another way of characterizing this fluctuating steady state
consists in looking at the probabilities $p(\C)$
of all the microscopic configurations $\C$.
Figure~\ref{figigs} shows a plot of the normalized probabilities
$2^{12}\,p(\C)$,
against the $2^{12}=4096$ configurations $\C$
of a column of 12 grains for $\eps=1$,
sorted according to lexicographical order (read down the column).
This plot exhibits a rugged and seemingly fractal structure
with self-similarity at all scales.
The most frequently visited configurations are
the $2^6=64$ degenerate ground states of the system, shown as open circles.

The knowledge of all the steady-state probabilities $p(\C)$
gives access to the entropy of the fluctuating steady state,
defined by means of the Boltzmann formula
\beq
S=-\sum_\C p(\C)\ln p(\C).
\eeq

\begin{figure}
\begin{center}
\includegraphics[angle=90,width=.75\linewidth]{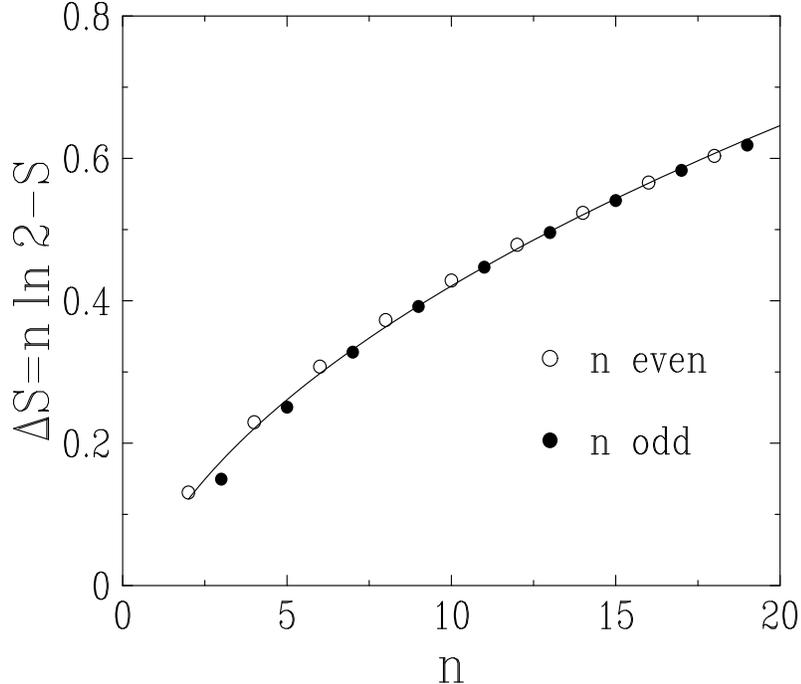}
\caption{\small
Plot of the entropy reduction $\Delta S=S_\flat-S$
in the zero-temperature steady state with $\eps=1$, against depth $n$.
Symbols: numerical data.
Full line: fit $\Delta S=(62\ln n+53)10^{-3}\,n^{1/3}$
(after~\cite{IV}).}
\label{figh}
\end{center}
\end{figure}

An estimate for the entropy $S$ can be derived
by using the main feature of the zero-temperature steady state,
i.e., the roughening law~(\ref{rough}).
Think again of the depth $n$ as a fictitious discrete time,
and of the local field~$h_n$ as the position of a random walker at time $n$.
For a free lattice random walk of $n$ steps,
one has $\mean{h_n^2}=n$, and the entropy reads $S_\flat=n\ln 2$,
as all configurations are equally probable.
In the present situation,
the entropy~$S$ is reduced with respect to $S_\flat$,
because $\mean{h_n^2}=W_n^2\ll n$.
The entropy reduction can be estimated as
$\Delta S=S_\flat-S\sim\sum_{m=1}^n1/W_m^2\sim n^{1/3}$.
Figure~\ref{figh} shows a plot of the entropy reduction $\Delta S$ against $n$.
A reasonable agreement with the above estimate is found.

The total entropy $S$ of the fluctuating steady state is therefore equal to
$S_\flat=n\ln 2$, up to a small subextensive reduction $\Delta S$
of order $n^{1/3}$.
The equality of the extensive part of $S$ with $S_\flat$
can be viewed as a manifestation
of the flatness hypothesis of Edwards~\cite{sam}.

\section{Low-temperature dynamics}

We now turn to the investigation of the dynamics of the model
for a low but non-zero temperature~$T$.
We consider for simplicity the case of an infinite $\xidy$.

If the slope $\eps$ is rational,
the rule~(\ref{zerodyrat}) is already stochastic at zero temperature,
so that no qualitatively novel effect appears at low temperature.

We therefore focus our attention onto the case of an irrational slope $\eps$.
We recall that the zero-temperature dynamics~(\ref{zerody})
drives the system to its quasiperiodic ground state,
where each orientation is aligned with its local field,
according to~(\ref{zerost}).
For a low but non-zero temperature $T$, there will be {\it mistakes},
i.e., orientations $\s_n=-\sign{ h_n}$ not aligned with their local field.
Equation~(\ref{zero}) suggests that the probability
of observing a mistake at site $n$ scales as
\beq
\Pi(n)\sim\exp\left(-\frac{2\abs{h_n}}{T}\right).
\eeq
Hence the most fragile sites $n$,
such that the local field $h_n$ is relatively small
in the ground state $(\abs{ h_n}\sim T\ll1)$,
will be preferred nucleation sites for mistakes,
and thus govern the low-temperature dynamics.

These nucleation sites can be located as follows.
Equation~(\ref{rot}) shows that the local field $h_n$ is small
when $n\Omega$ is close to an integer $m$.
The latter turns out to be $m=m^+_n$.
Indeed
\beq
n\Omega=m+\delta\Longrightarrow h_n=\frac{\delta}{1-\Omega},
\qquad(\Omega-1<\delta<\Omega).
\eeq
The most active nucleation sites are therefore in correspondence
with rational numbers $m/n$ which are the closest
to the irrational rotation number~$\Omega$.
Finding these rational approximants is a well-defined problem of Number Theory,
referred to as Diophantine approximation~\cite{hr}.

Let us illustrate this on the example
of the golden-mean slope~(\ref{gold}).
We are led to introduce the Fibonacci numbers
$F_k$~\cite{quasi,hr}, defined by the recursion formula
\beq
F_k=F_{k-1}+F_{k-2}\qquad(F_0=0,\qquad F_1=1).
\eeq
We have alternatively
\beq
F_k=\frac{\Phi^k-(-\Phi)^{-k}}{\sqrt5}.
\eeq
The leading nucleation sites are the Fibonacci sites $n=F_k$.
We have $m=m^+_n=F_{k-2}$, $m^-_n=F_{k-1}$, and $h_n=(-)^k\,\Phi^{-(k-1)}$,
so that
\beq
\Pi_k=\Pi(F_k)\sim\exp\left(-\frac{2\Phi}{\sqrt5\,T F_k}\right).
\label{pik}
\eeq

We can therefore draw the following picture of low-temperature dynamics.
Mistakes are nucleated at the Fibonacci sites $F_k$,
according to a Poisson process,
with exponentially small rates proportional to the $\Pi_k$.
They are then advected with constant velocity $V\approx2.58$,
just as in the zero-temperature case.
The system is ordered according to its quasiperiodic ground state
in an upper layer $(n<\N(t))$, while the rest is disordered,
somehow like the zero-temperature steady state for a rational slope.
The depth $\N(t)$ of the ordered layer,
given by the position of the uppermost mistake,
is therefore a natural collective co-ordinate
describing low-temperature dynamics.

\begin{figure}
\begin{center}
\includegraphics[angle=90,width=.75\linewidth]{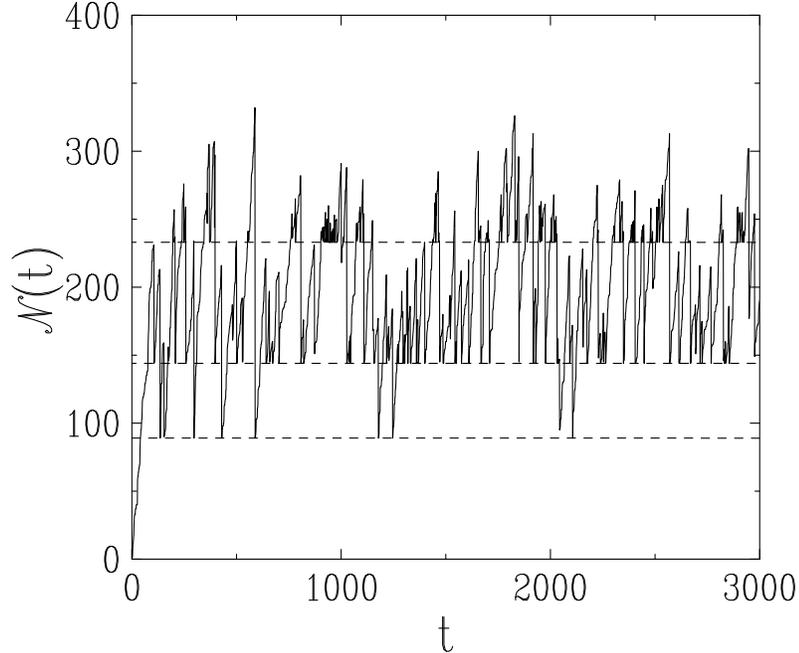}
\end{center}
\caption{
Plot of the instantaneous depth $\N(t)$ of the ordered layer
against time $t$, for the golden-mean slope, and temperature $T=0.003$.
Dashed lines: leading nucleation sites given by consecutive Fibonacci numbers
(bottom to top: $F_{11}=89$, $F_{12}=144$, $F_{13}=233$)
(after~\cite{III}).}
\label{figk}
\end{figure}

Figure~\ref{figk} shows a typical sawtooth plot of the instantaneous
depth $\N(t)$, for a temperature $T=0.003$.
The leading nucleation sites are observed to be given by Fibonacci numbers.

The system thus reaches a steady state,
characterized by a finite ordering length $\mean{\N}$,
which diverges at low temperature, as mistakes become more and more rare.
The law of this divergence can be predicted by the following argument.
The most active nucleation Fibonacci site is such that
the nucleation time $1/\Pi_k$ is comparable to the advection time
to the next nucleation site $F_{k+1}$,
i.e., $(F_{k+1}-F_k)/V\approx F_k/(\Phi V)$, hence $\Pi_k F_k\sim\Phi V$.
Less deep sites have too small nucleation rates,
while the mistakes nucleated at deeper sites have little chance to be
the uppermost ones.
Equations~(\ref{pik}) then yields
\beq
\frac{\sqrt5}{2\Phi}\,T F_k\ln\frac{F_k}{\Phi V}\approx1,
\label{fklnfk}
\eeq
hence
\beq
\mean{\N}\approx F_k\approx\frac{2\Phi}{\sqrt5}\,\frac{\absln T}{T}.
\eeq
The estimate~(\ref{fklnfk}) correctly predicts that $F_{12}=144$
is the most active nucleation site at temperature~$T=0.003$
(see Figure~\ref{figk}).

Both the existence of a well-defined sequence of preferred nucleation sites,
and the divergence of the depth of the most active nucleation site
as $\mean{\N}\sim\absln T/T$,
are very reminiscent of the phenomenon of hierarchical melting,
put forward by Serge Aubry and collaborators~\cite{hime}.
This effect manifests itself as peaks in the specific heat at low temperature
in some incommensurate modulated solids,
each peak being in correspondence with a temperature-dependent spatial scale,
where the system is more fragile and preferentially melts.

\section{Outline}

In a series of recent works ~\cite{I,II,III,IV}
we have introduced and investigated in detail
a columnar {\it toy model} for granular compaction.
The minimal model of interest described in this report
has one essential parameter, the shape parameter $\eps$,
which gives the slope of the geometrical construction of ground states
within the superspace formalism.

Rational and irrational values of $\eps$ correspond to very different
kinds of behavior, both in statics (structure of ground states)
and in low-temperature dynamics (retrieval of these ground states).

{\bf For irrational values of $\eps$}:
\begin{itemize}
\item
The model has a unique quasiperiodic ground state.
\item
Zero-temperature dynamics leads to a fast retrieval of this ground state
by means of the ballistic growth of the thickness of the ordered top layer.
\item
A crossover to a logarithmic growth takes place when this depth
becomes comparable to the dynamical length.
\item
Low-temperature dynamics involves the nucleation of mistakes
at preferred nucleation sites.
This phenomenon is analogous to hierarchical melting.
\end{itemize}

{\bf For rational values $\eps=p/q$}:
\begin{itemize}
\item
The model has extensively degenerate ground states
and therefore a positive configurational entropy.
\item
Ground states consist of random sequences of two well-defined patterns
consisting of $p+q$ grains.
\item
Zero-temperature dynamics is not able to retrieve any of the ground states.
\item
The system is rather driven into a non-trivial fluctuating steady state
characterized by unbounded,
albeit subextensive fluctuations in the local fields.
\end{itemize}


\begin{thebibliography}{99}

\bibitem{I}
P.F. Stadler, A. Mehta, and J.M. Luck, Adv. Complex Systems {\bf 4}, 429 (2001).

\bibitem{II}
P.F. Stadler, J.M. Luck, and A. Mehta, Europhys. Lett. {\bf 57}, 46 (2002).

\bibitem{III}
A. Mehta and J.M. Luck, J. Phys. A {\bf 36}, L365 (2003).

\bibitem{IV}
J.M. Luck and A. Mehta, Eur. Phys. J. B {\bf 35}, 399 (2003).

\bibitem{deB}
N.G. de Bruijn, Nederl. Akad. Wetens. Proc. A {\bf 84}, 27 (1981).

\bibitem{quasi}
M. Duneau and A. Katz, Phys. Rev. Lett. {\bf 54}, 2688 (1985).
\\
M. Duneau and A. Katz, J. Phys. (France) {\bf 47}, 181 (1986).
\\
V. Elser, Phys. Rev. B {\bf 32}, 4892 (1985).
\\
P.A. Kalugin, A.Yu. Kitayev, and L.S. Levitov, JETP Lett. {\bf 41}, 145 (1985).
\\
P.A. Kalugin, A.Yu. Kitayev, and L.S. Levitov, J. Phys. (France) Lett.
{\bf 46}, L601 (1985).

\bibitem{hr}
G.H. Hardy and E.M. Wright, {\it An Introduction to the Theory of Numbers}
(Clarendon, Oxford, 1990).

\bibitem{sconf}
J. J\"ackle, Phil. Mag. B {\bf 44}, 533 (1981).
\\
R.G. Palmer, Adv. Phys. {\bf 31}, 669 (1982).

\bibitem{bray}
A.J. Bray, Adv. Phys. {\bf 43}, 357 (1994).

\bibitem{sam}
S.F. Edwards, in {\it Granular Matter: An Interdisciplinary Approach}, ed. A.
Mehta (Springer, New York, 1994).

\bibitem{hime}
F. Vallet, R. Schilling, and S. Aubry, Europhys. Lett. {\bf 2}, 815 (1986).
\\
R. Schilling and S. Aubry, J. Phys. C {\bf 20}, 4881 (1987).
\\
F. Vallet, R. Schilling, and S. Aubry, J. Phys. C {\bf 21}, 67 (1988).

\end{thebibliography}
\end{document}